\documentclass[%
 reprint,
nofootinbib,
 amsmath,amssymb,
 aps,
]{revtex4-1}

\usepackage{graphicx, fancybox, color}
\usepackage{amsmath,amssymb,bm}
\usepackage{braket}
\newcommand{\bvec}[1]{\mbox{\boldmath $#1$}}
\newcommand{\QQ}{Q\bar{Q}}

\begin{document}

\title{
Dynamical dissociation of quarkonia by wave function decoherence
}

\author{Shiori Kajimoto}
\email{kajimoto@kern.phys.sci.osaka-u.ac.jp}
\affiliation{%
 Department of Physics, Osaka University, Toyonaka, Osaka 560-0043, Japan
}%
\author{Yukinao Akamatsu}%
\email{akamatsu@kern.phys.sci.osaka-u.ac.jp}
\affiliation{%
 Department of Physics, Osaka University, Toyonaka, Osaka 560-0043, Japan
}%
\author{Masayuki Asakawa}%
\email{yuki@phys.sci.osaka-u.ac.jp}
\affiliation{%
 Department of Physics, Osaka University, Toyonaka, Osaka 560-0043, Japan
}%
\author{Alexander Rothkopf}
\email{rothkopf@thphys.uni-heidelberg.de}
\affiliation{%
Institute for Theoretical Physics,
Department of Physics and Astronomy,
Heidelberg University,
D-Heidelberg 69120, Germany
}

\begin{abstract}
We investigate the real-time evolution of quarkonium bound states in a quark-gluon plasma in one dimension using an improved QCD-based stochastic potential model.
This model describes the quarkonium dynamics in terms of a Schr\"odinger equation with an in-medium potential and two noise terms encoding the residual interactions between the heavy quarks and the medium.
The probabilities of bound states in a static medium and in a boost-invariantly expanding quark-gluon plasma are discussed.
We draw two conclusions from our results:
One is that the outcome of the stochastic potential model is qualitatively consistent with the experimental data in relativistic heavy-ion collisions.
The other is that the noise plays an important role in order to describe quarkonium dynamics in medium; in particular it causes decoherence of the quarkonium wave function.
The effectiveness of decoherence is controlled by a new length scale $l_{\rm corr}$.
It represents the noise correlation length and its effect has not been included in existing phenomenological studies.
\end{abstract}

\date{\today}

\maketitle

\section{\label{sec:level1}Introduction}
Ultrarelativistic heavy-ion collisions are the only experiments currently able to create the high-temperature state of nuclear matter on the Earth.
By colliding two nuclei accelerated to almost the speed of light, the temperature in the small collision volume of nuclear size $\sim (10 \ {\rm fm})^3$ reaches above 2 trillion kelvin (about 0.2 GeV) for a short period of time $\sim$ 10 fm/{\it c}.
In such an extremely hot environment, nuclear matter is expected to take on the form of a strongly coupled plasma, composed of quarks and gluons [quark-gluon plasma (QGP) \cite{Yagi:985487}], liberated from inside the
nucleons.
Evidence of QGP formation in heavy-ion collisions at the Relativistic Heavy Ion Collider (RHIC) and the Large Hadron Collider (LHC) is being accumulated by combining various indirect hadronic and leptonic signals \cite{Jacak:2012dx}.

Quarkonium is a bound state of a heavy quark pair ($c\bar c$ charmonium, $b\bar b$ bottomonium) and is an ideal probe to signal the QGP formation \cite{Aarts:2016hap}.
The binding force of the heavy quark pair in the vacuum is strong and reaches over long distance.
In QGP, however, the liberated quarks and gluons screen the charges of heavy quarks and the binding force gets short ranged.
This led to the idea  that the heavy quark pair in QGP cannot be bound by the weakened force and the number of quarkonium states would decrease inside QGP \cite{Matsui:1986dk}.
Once a quarkonium state cannot maintain itself, the heavy quarks diffuse independently in the QGP.
It is expected that when the heavy quark density is high, though, there is yet a finite probability that initially uncorrelated heavy quarks form a quarkonium state at the freeze-out of QGP fireball via recombination \cite{BraunMunzinger:2000px}.
%Therefore, depending on the heavy quark density and the quarkonium dissociation rate, the QGP formation does not necessarily result in the suppression of quarkonium yield.
%Roughly, the density of heavy quarks with smaller mass are higher and shallower bound states dissociate more easily. 
%Thus, charmonia yield tends to receive contamination from recombination, especially in heavy-ion collisions at higher energies.

Experimental yields of $\Upsilon (b\bar b)$ in Pb-Pb collisions at the LHC \cite{Chatrchyan:2011pe,Chatrchyan:2012lxa,Khachatryan:2016xxp,Abelev:2014nua} and of $\Upsilon$ \cite{Adamczyk:2013poh,Adare:2014hje} and $J/\psi (c\bar c)$
\cite{Adare:2006ns,Adare:2011yf,Adamczyk:2013tvk} in Au-Au collisions at the RHIC indeed show suppression compared to that in proton-proton collisions scaled with the number of binary nucleon-nucleon collisions.
Its magnitude is more or less consistent with in-medium dissociation of the quarkonia \cite{Strickland:2011mw,Strickland:2011aa,Krouppa:2015yoa,Krouppa:2016jcl,Song:2011nu,Emerick:2011xu,Zhou:2014hwa,Rapp:2008tf, Sharma:2012dy, Aronson:2017ymv}.
Interestingly, the yield of $J/\psi$ in Pb-Pb collisions at the LHC is less suppressed than that in Au-Au collisions at the RHIC \cite{Abelev:2013ila,Adam:2016rdg}.
Analysis of charm quark chemical reactions in the QGP suggests that recombination of initially uncorrelated charm and anticharm quarks does play an important role for $J/\psi$ production in heavy-ion collisions \cite{Zhao:2010nk,Zhao:2011cv,Zhou:2014kka}.

Recently rapid theoretical progress has been made in the quantum mechanical description of heavy quarks in a QGP.
A potential model picture for a heavy quark pair matches our physics intuition even in a finite temperature environment. 
However, in a thermal environment, the concept of potential itself is not obvious.
The heavy quark potential in the QGP is derived by properly integrating out the QGP degrees of freedom.
An important observation was first made in analyses of the thermal Wilson loop that an imaginary part appears in the heavy quark potential because of the scattering between the heavy quarks and the plasma constituents \cite{Laine:2006ns,Beraudo:2007ky,Brambilla:2008cx}.
There are ongoing efforts in order to numerically compute this quantity from lattice QCD simulations \cite{Rothkopf:2011db,Burnier:2014ssa,Burnier:2016mxc,Petreczky:2017aiz}. 
Then a dynamical implementation of this complex potential in the context of the Schr\"odinger equation was first given by means of stochastic potential in which thermal noise is added to a real valued potential \cite{Akamatsu:2011se}.
The thermal noise in the stochastic potential describes an effective coupling between the heavy quarks and the plasma constituents.
In a broader perspective, the complex potential and its dynamical implementation by a stochastic potential can be consistently derived from QCD using the framework of open quantum systems \cite{Akamatsu:2014qsa,BRE02}. 
Applications of the theory of open quantum systems to heavy quark physics in the QGP have also appeared in 
\cite{Young:2010jq,Borghini:2011ms,Blaizot:2015hya,Akamatsu:2015kaa,Brambilla:2016wgg,DeBoni:2017ocl, Katz:2015qja}.

In this paper, we study how a quarkonium evolves in such a stochastic potential.
The noise in the stochastic potential has a finite correlation length $l_{\rm corr}$ reflecting the finite momentum transfer in a scattering process between a heavy quark and a medium particle \cite{Akamatsu:2011se}.
Perturbative calculations show that the noise correlation length is of the order of the screening length \cite{Akamatsu:2014qsa}.
An important effect of such a noise is the appearance of wave function decoherence.
When the size of the quarkonium wave function is larger than the correlation length, the wave function is disturbed by uncorrelated noise and its coherence is easily lost.
On the other hand, when the quarkonium wave function is localized in a smaller size than the correlation length, it remains almost undisturbed.
In a realistic event, quarkonia with different extent of wave functions are present in a QGP fireball with continuously decreasing temperature.
To obtain an essential insight into these intricate dynamics, we numerically investigate the effect of decoherence caused by the stochastic potential in a simplified one-dimensional model of a static, as well as a boost-invariantly expanding QGP.

The organization of this paper is as follows. 
In Sec.~\ref{sec:level2}, we introduce the improved stochastic potential model, which describes quarkonium dynamics in QGP.
To illustrate the difference between the stochastic potential and the complex potential, we subsequently discuss the decay rate of quarkonium states.
In Sec.~\ref{sec:level3}, we present our numerical results in one dimension.
The quarkonium evolution in a QGP with a fixed temperature is studied in Sec.~\ref{subsec:level3_1} and that in a boost-invariantly expanding QGP with decreasing temperature is discussed in Sec.~\ref{subsec:level3_2}.
We summarize our analysis in Sec.~\ref{sec:level4}.

\section{\label{sec:level2}Stochastic Potential Model}

First let us introduce the state-of-the art formulation of the stochastic potential as derived from QCD \cite{Akamatsu:2014qsa}.
Suppose that a heavy quark is located at $\bvec{x}\equiv \bvec{R}+\bvec{r}/2$ and a heavy antiquark at $\bvec{y}\equiv\bvec{R}-\bvec{r}/2$ in a hot QGP medium. 
They are exposed to thermal fluctuations and the strength of the potential between them is not only weakened by screening but also fluctuates. 
These two effects are described by a screened potential $V(\bm r)$ and by a stochastic term $\Theta(\bm r,t)$ in the Hamiltonian for their relative motion\footnote{
Because it is based on the Hamilton formalism, the stochastic potential cannot describe dissipation which becomes important when the system approaches to equilibrium in a longer time scale.
},
\begin{align}
\nonumber H(\bm r, t) &\equiv  -\nabla^2_{\bm r}/M+V(\bm r) +\Theta(\bvec{r},t), \\[2ex]
\Theta(\bvec{r},t) &\equiv \theta(\bvec{R}+\bvec{r}/2,t)-\theta(\bvec{R}-\bvec{r}/2,t).
\end{align}
Here $M$ is the heavy (anti)quark mass, and $\Theta$ is the sum of noise terms for the heavy quark ($\theta(\bvec{x},t)$) and antiquark ($-\theta(\bvec{y},t)$).
The microscopic origin of $\theta$ is the interaction between the heavy (anti)quark and plasma constituents and it models the effect of thermal fluctuations which have been already integrated when deriving the in-medium potential.
The noise terms for the heavy quark and antiquark in $ \Theta(\bvec{r},t)$ have different signs because they represent opposite color charges.
$\theta$ represents a Gaussian white noise with finite correlation length: 
\begin{subequations}
\begin{align}
&\langle\theta(\bvec{x},t)\rangle=0, \\[2ex]
&\langle\theta(\bvec{x},t)\theta(\bvec{x'},t')\rangle= D(\bvec{x}-\bvec{x'})\delta(t-t'),
\end{align}
\end{subequations}
from which follows
\begin{align}
\langle\Theta(\bm r,t)\Theta(\bm r',t')\rangle
=&2\left[D\left(\frac{\bm r-\bm r'}{2}\right)-D\left(\frac{\bm r+\bm r'}{2}\right)\right] \nonumber \\[2ex]
&\times\delta(t-t') \,.
\end{align}
Note that in previous works \cite{Akamatsu:2011se,Rothkopf:2013kya} the expression for the noise $\Theta$ was modeled based on intuition and not systematically derived. 
In the more recent systematic derivation of the open quantum system approach to quarkonia from QCD \cite{Akamatsu:2014qsa}, it has been found that the noise $\Theta$ has to be constructed from two noises for a heavy quark and antiquark.

Once time is discretized, the delta function $\delta(t-t')$ is expressed as $\frac{\delta_{t t'}}{\Delta t}$.
Taking $\Delta t\to 0$, the noise scales as $(\Delta t)^{-1/2}$.
The unitary evolution operator $e^{-i H\Delta t}$ for the stochastic potential model naturally explains the complex potential found in \cite{Laine:2006ns,Beraudo:2007ky,Brambilla:2008cx} while conserving the wave function norm.
The operator is expanded in terms of the infinitesimal time $\Delta t$,
\begin{align}
\label{eq:time evolution}
e^{-i\Delta t H} 
&\simeq 1-i\Delta t H(\bm r, t) -\frac{1}{2}\left(\Delta t\Theta(\bm r,t)\right)^2 + \mathcal O(\Delta t^{3/2}) \nonumber \\[2ex]
&\equiv 1 -i\Delta t H_{\rm eff}(\bm r, t),
\end{align}
and a stochastic Schr\"odinger equation for a quarkonium wave function $\Psi_{\QQ}$ is obtained,
\begin{align}
\label{eq:sse}
i \frac{\partial}{\partial t}\Psi_{\QQ}(\bvec{r},t) = H_{\rm eff}(\bm r, t) \Psi_{\QQ}(\bvec{r},t).
\end{align}
Using $H_{\rm eff}$, the evolution equation for the noise averaged quarkonium wave function $\langle\Psi_{Q\bar Q}\rangle$ can be written as a Schr\"{o}dinger equation with a complex potential,
\begin{align}
i \frac{\partial}{\partial t}&\langle\Psi_{\QQ}(\bvec{r},t)\rangle = \langle H_{\rm eff}(\bm r, t)\rangle\langle\Psi_{\QQ}(\bvec{r},t)\rangle \\[2ex]
&=\left[-\nabla^2_{\bm r}/M+V(\bm r)-i\{D(\bvec{0})-D(\bvec{r})\} \right]\langle\Psi_{\QQ}(\bvec{r},t)\rangle. \nonumber
\end{align}
Note that the complex potential in the literature is actually defined using the time evolution of $\langle\Psi_{Q\bar Q}\rangle$ (not $\Psi_{Q\bar Q}$) in the $M\to\infty$ limit and its imaginary part does not indicate the violation of unitarity.

By comparing the generators of the time evolution for $\langle\Psi_{Q\bar Q}\rangle$, we can match the stochastic potential model with the underlying microscopic theory of QCD\footnote{
In leading order perturbation theory, a quantum master equation for the reduced density matrix of a quarkonium state may be obtained, whose recoilless limit corresponds to the stochastic potential model \cite{Akamatsu:2014qsa}.The parameters of the stochastic potential obtained in this fashion are identical to those obtained by matching to the perturbative complex potential below.
}.
From the stochastic potential model in the $M\to\infty$ limit, the generator reads
\begin{align}
\lim_{M\to \infty}\langle H_{\rm eff}(\bm r, t)\rangle = V(\bm r) -i\{D(\bvec{0})-D(\bvec{r})\},
\end{align}
while that from microscopic theory is in general expressed by a complex potential $V_{\rm Re}(\bm r) + iV_{\rm Im}(\bm r)$ after taking the Markovian limit and the matching conditions are
\begin{subequations}
\begin{align}
V_{\rm Re}(\bm r) &= V(\bm r), \\[2ex]
V_{\rm Im}(\bm r) &= D(\bvec{r})-D(\bvec{0}).
\end{align}
\end{subequations}
We can immediately see that $V_{\rm Im}(\bm r=\bm 0)=0$ must be satisfied to be consistent with the stochastic potential.
It is natural to assume that the noise correlation vanishes when separated by a long distance,
\begin{align}
\lim_{r\to \infty}D(\bm r)=0;
\end{align} 
therefore by matching, we get
\begin{align}
D(\bm r) = V_{\rm Im}(\bm r) - \lim_{r\to\infty}V_{\rm Im}(\bm r).
\end{align}
In leading order perturbative calculations for a QGP with $N_c$ colors and $N_f$ flavors of massless quarks, the complex potential is obtained as \cite{Laine:2006ns,Beraudo:2007ky,Brambilla:2008cx}
\begin{subequations}
\label{eq:Lainepot}
\begin{align}
V_{\rm Re}(\bm r)&=- \frac{g^2 C_F}{4 \pi}  \left[ m_D + \frac{\exp(-m_D r)}{r} \right], \\
V_{\rm Im}(\bm r)&=- \frac{ g^2 T C_F}{4 \pi} \phi (m_D r),
\end{align}
\end{subequations}
where $C_F=(N_c^2-1)/2N_c$, $m_D^2=(N_c/3+N_f/6)g^2T^2$, and $\phi(x)$ is a monotonically increasing function of $x$:
\begin{align}
\phi(x)&= 2 \int^{\infty}_0 \frac{dz \ z}{(z^2 +1)^2} \left[ 1- \frac{\sin(zx)}{zx} \right],
\end{align}
with $\phi(0)=0$ and $\phi(\infty)=1$.
Here, a scale hierarchy $T\gg 1/r\agt m_D$ is assumed.
The complex potential above implies for the stochastic potential having the following two features:
(i) Color charges are screened with a screening length $1/m_D$ and
(ii) the noise $\theta$ is correlated over a correlation length $\sim 1/m_D$.
The latter leads to a new dynamical effect on a quarkonium: Wave function decoherence.
It should be emphasized here that a stochastic potential itself is a generic notion underlying the in-medium complex potential and is not restricted to the above scale hierarchy.
In this paper, we adopt Eq.~\eqref{eq:Lainepot} as a guideline to model the in-medium potential even at temperatures close to the deconfinement transition.

The noise correlation length $l_{\rm corr}\sim 1/m_D$ introduces a new scale for a quarkonium with a coherence length $l_{\Psi}$.
When the wave function is localized $l_{\Psi}\ll l_{\rm corr}$, the noise cannot recognize that there are two opposite charges with separation $l_{\Psi}$ and the wave function is undisturbed.
When the wave function is extended over a long distance $l_{\Psi}\agt l_{\rm corr}$, the noise recognizes the heavy quark and antiquark and kicks them incoherently.
In the former case the quarkonium state remains virtually unchanged, while in the latter case it is easily mixed with excited states.
Therefore, in addition to the static effects (such as melting) by a screened potential, decoherence provides another dynamical mechanism for quarkonium suppression.
We see that in practice an intricate interplay between screening and decoherence ensues, which also entails the thermal excitation and deexcitation of states, all of which modify the final yield of the individual bound states.

Let us comment on the difference between evolving heavy quarkonium with a stochastic potential \cite{Akamatsu:2011se} and, as has been done in the literature so far, with a complex potential \cite{Strickland:2011mw,Strickland:2011aa}.
To be specific, we compare the decay rates of an initial state computed according to these two approaches.
We take the initial state to be an eigenstate $\varphi_n$ of the Hamiltonian without noise,
\begin{align}
\left[-\nabla_{\bm r}^2/M + V(r)\right]\varphi_n(\bm r) = E_n\varphi_n(\bm r).
\end{align}
In a single time step, the stochastic potential evolves $\varphi_n$ into $\varphi(\Delta t)$ via
\begin{align}
\varphi(\bm r,\Delta t) = \left[1-i\Delta t H_{\rm eff}(\bm r, t)\right]\varphi_n(\bm r),
\end{align}
and the corresponding occupation $c_n(\Delta t)$ is given by
\begin{align}
c_n(\Delta t)=&\Bigl\langle\left|\int_{\bm r}\varphi_n(\bm r)^*\varphi(\Delta t, \bm r)\right|^2\Bigr\rangle \\[2ex]
\simeq &1-2\Delta t\int_{\bm r}\left(D(\bm 0)-D(\bm r)\right)|\varphi_n(\bm r)|^2 \nonumber \\[2ex]
&+(\Delta t)^2\int_{\bm r, \bm r'}\langle\Theta(\bm r,0)\Theta(\bm r',0)\rangle|\varphi_n(\bm r)|^2|\varphi_n(\bm r')|^2. \nonumber 
\end{align}
The decay rate $\Gamma_n\equiv -\frac{dc_n}{dt}\bigr|_{t=0}$ is obtained as
\begin{align}
\label{eq:decay}
\Gamma_n
=&2\int_{\bm r}\left[D(\bm 0)-D(\bm r)\right]|\varphi_n(\bm r)|^2 \\[2ex]
&-2\int_{\bm r,\bm r'} \left[D\left(\frac{\bm r-\bm r'}{2}\right)-D\left(\frac{\bm r+\bm r'}{2}\right)\right]\nonumber \\[2ex]
& \hspace{4em} \times |\varphi_n(\bm r)|^2 |\varphi_n(\bm r')|^2.\nonumber 
\end{align}
The second term of the right-hand side in Eq.~\eqref{eq:decay} originates from the correlations within the noise.
Using the complex potential, the decay rate is given only by the first term.
%For eigenfunctions with definite parity and {\red for two equal masses},
The second term vanishes due to the parity of eigenstates and the two approaches give the same decay rate \footnote{
When the system is composed of a heavy quark and antiquark with different masses ($M_1$ and $M_2$), they are located at $\bm R+s\bm r$ and $\bm R-(1-s)\bm r$ with $s\equiv M_1/(M_1+M_2)$.
In this case the decay rates are always different in the two approaches.
}.
Even though the evolution starts with a common decay rate, the wave function evolves differently in the two approaches.
As we have seen, a wave function evolved with the complex potential corresponds to an averaged wave function evolved with the stochastic potential.
Therefore the occupation probability of any state is always {\it underestimated} in the complex potential.

\section{\label{sec:level3}One-dimentional Numerical Calculation}

We solve the stochastic Schr\"odinger equation \eqref{eq:sse} numerically and compute the survival probability of each state.
To demonstrate the effect of decoherence on quarkonium bound states, it is sufficient to simulate a one-dimensional system.
We first consider a static one-dimensional system in Sec.~\ref{subsec:level3_1} and subsequently a Bjorken expanding system with decreasing temperature in Sec.~\ref{subsec:level3_2}.

The numerical calculation is performed using an operator splitting method.
The evolution operator from time $t_n$ to $t_n+\Delta t$ is given by
\begin{align}
U_{t_n\to t_n+\Delta t}&=e^{-i\Delta t H}\nonumber \\[2ex]
&\simeq U^{\Theta}_{t_n\to t_n+\Delta t} U^{\langle H\rangle}_{t_n\to t_n+\Delta t} + \mathcal O(\Delta t^{3/2}),
\end{align}
where $U^{\Theta}_{t_n\to t_n+\Delta t}$ is a random phase rotation with spatially correlated noise:
\begin{align}
U^{\Theta}_{t_n\to t_n+\Delta t} = \exp\left[-i\Delta t\Theta(x,t_n)\right],
\end{align}
and $U^{\langle H\rangle}_{t_n\to t_n+\Delta t}$ evolves the wave function with the Hamiltonian without noise $\langle H\rangle$ (not $\langle H_{\rm eff}\rangle$) using the Crank-Nicolson scheme.
Both $U^{\Theta}_{t_n\to t_n+\Delta t}$ and $U^{\langle H\rangle}_{t_n\to t_n+\Delta t}$ are manifestly unitary.
In our numerical implementation for bottomonium (charmonium), the one-dimensional spatial axis $-2.56 {\rm fm} \leq x \leq 2.56 {\rm fm}$ ($-5.12 {\rm fm} \leq x \leq 5.12 {\rm fm}$) is discretized with 512 cells (1024 cells) of size $\Delta x = 0.01$ fm and the wave function is updated 100000 times with a time step $\Delta t =0.0001$ fm from $t=0$ to $t=10$ fm.
The spatial size of 5.12 fm (10.24 fm) is large enough to accommodate bound state wave functions in our computations.
We collect 1000 events and take their average to produce the thermal ensemble average for each setup.

To simplify the modeling, we parametrize $V(x)$ and $D(x)$ in such a way that their essential features are captured:
\begin{subequations}
\begin{align}
\label{eq:stoch_pot_model}
V(x) &= -\frac{\alpha_{\rm eff}}{r}\exp\left(-m_D |x|\right),\\[2ex]
D(x) &= \gamma \exp\left(-|x|^2/l_{\rm corr}^2\right),
\end{align}
\end{subequations}
with parameters $\alpha_{\rm eff}$, $m_D$, $\gamma$, and $l_{\rm corr}$.
The values for these parameters obtained in a perturbative calculation are
\begin{subequations}
\begin{align}
\alpha_{\rm eff} &= \frac{g^2 C_F}{4\pi}, \ \ m_D = \sqrt{\frac{N_c}{3}+\frac{N_f}{6}}gT, \\[2ex]
\gamma  &= \frac{g^2 T C_F}{4\pi}, \ \ l_{\rm corr} \sim \frac{1}{gT}.
\end{align}
\end{subequations}
We only evaluate the scaling with $g$ and $T$ for the noise correlation length $l_{\rm corr}$ because the noise correlation $D(x)$ is not exactly a Gaussian function.
In numerical calculations, the singularity of the Debye screened potential at the origin $x=0$ needs to be regularized.
We define $\tilde x(x) \equiv {\rm sgn}(x)\sqrt{x^2 + 1/M^2}$,  with which the Debye screened potential is regularized as $V(x)\to V(\tilde x(x))$.

\subsection{\label{subsec:level3_1} Quarkonium in a static QGP}
\begin{table}[b]
\caption{Mass and parameters in the model}
\label{table:1}
\begin{center}
\begin{tabular}{ccccc} \hline \hline
$M$ [GeV]	& \  $\alpha_{\rm eff} \ $ 	& $m_D$ [GeV]	 	& \ $\gamma$ [GeV] \	& $l_{\rm corr} $ [fm]\\ \hline
4.8 			&0.3 					& 0.4				& $ 0.012$  			& 0.04-0.96 \\ \hline\hline
\end{tabular} 
\end{center}
\end{table}

We here compute the time evolution of a bottomonium wave function under the stochastic potential to investigate the effects of decoherence.
The mass of the bottom quark and the parameters of the stochastic potential are listed in Table~\ref{table:1}, which correspond to estimates for a QGP at $T=0.4$ GeV, a typical temperature in relativistic heavy-ion collisions.
We change the noise correlation length $l_{\rm corr} = 0.04, 0.16, 0.32, 0.48, 0.96$ fm to study how the results depend on $l_{\rm corr}$. When we take $l_{corr}\sim 1/T$, $l_{corr}$ is about $0.5$ fm. 

The initial wave function is chosen to be the ground state in the (regularized) Debye screened potential $V(\tilde x(x))$ in Eq.~\eqref{eq:stoch_pot_model}.
The radius, or the coherence length $l_{\Psi}$, of the ground state is about 0.2 fm.
The wave function is evolved by the stochastic Schr\"odinger equation \eqref{eq:sse}.
In each event, we compute the occupation probability of the ground state as a function of time, then take an ensemble average over 1000 events.
Note that with this setup, the change in the occupation probability is solely due to the noise term $\Theta$.

Our results are shown in Fig.~\ref{fig:fixedT}.
We can clearly see that the occupation probability of the ground state is sensitive to the noise correlation length $l_{\rm corr}$.
When $l_{\rm corr}$ is chosen at the upper end of the parameter range, the initial ground state is only weakly affected by the noise.
On the other hand, when a shorter $l_{\rm corr}$ is chosen, the initial ground state is easily excited by the noise and the wave function becomes a mixture of the ground and excited states.\footnote{
The dissipation might become relevant for shorter $l_{\rm corr}$ in a time scale of 9 fm, but it is beyond the scope of our analysis.
}
The transition of these two regimes roughly takes place at $l_{\rm corr}\sim l_{\Psi}\sim$ 0.2 fm.

\newlength{\bw}
\newsavebox{\boxa}
\sbox{\boxa}{\includegraphics[width=9cm,bb= 0 0 380 255,clip]{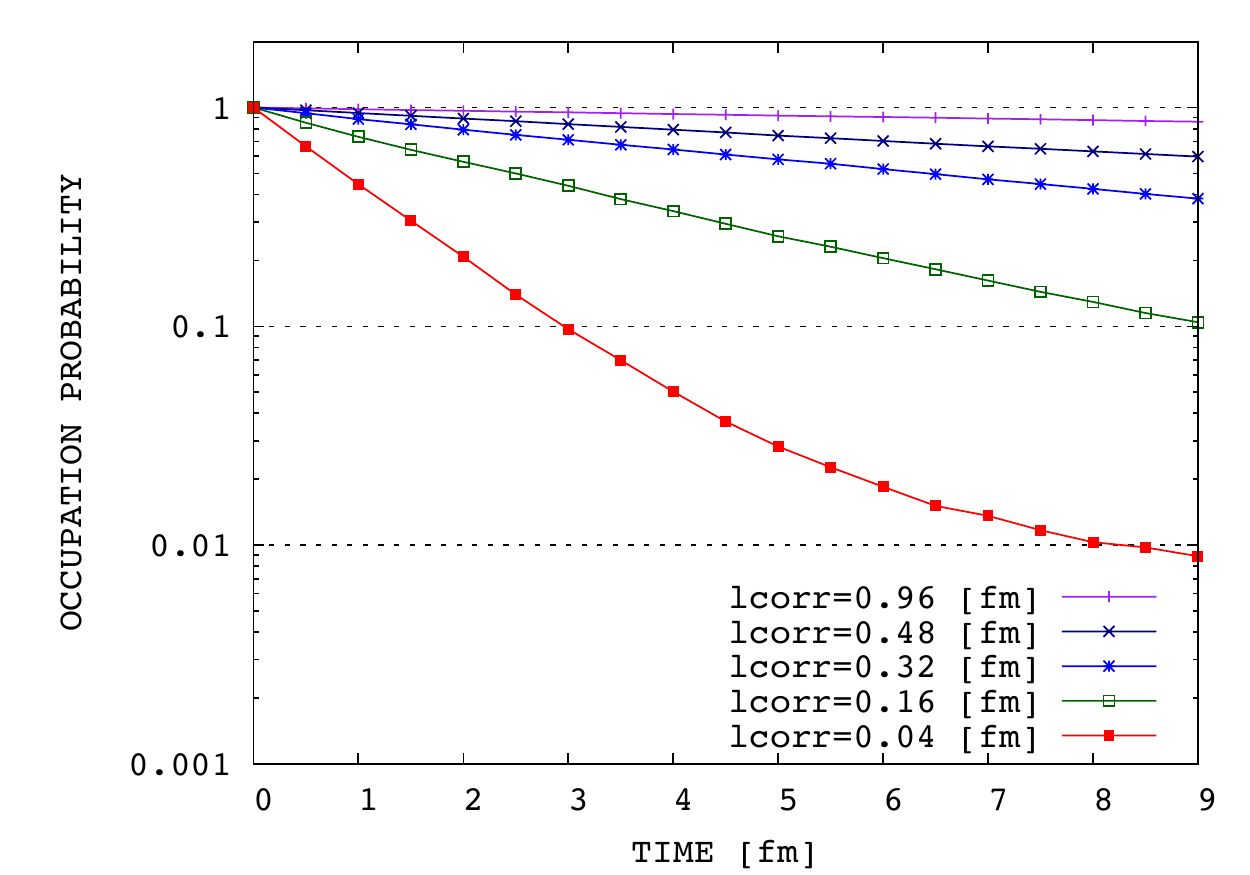}}
\settowidth{\bw}{\usebox{\boxa}}

\begin{figure}
\centering
\parbox{\bw}{\usebox{\boxa}} 
\caption{
Time evolution of the ground state  occupation probability in the stochastic potential model in a static QGP.
Both the initial state and the projected state are the ground state in the Debye screened potential.
The noise correlation length ranges from $l_{\rm corr}$=0.04 to 0.96 fm.
}
\label{fig:fixedT}
\end{figure}

Let us compare the evolution from the ground state using the stochastic potential and the complex potential as discussed in Sec.~\ref{sec:level2}.
We choose the noise correlation length  $l_{\rm corr}$=0.48, 0.16 fm.
Note that the complex potential in this computation corresponds to $V_{\rm Re}(x)=V(\tilde x(x))$ and $V_{\rm Im}(x)=D(x)-D(0)$ in Eq.~\eqref{eq:stoch_pot_model}.
The computation of the time evolution of the wave function under the complex potential is carried out via the following procedure:
(i) The wave function is evolved by the stochastic Schr\"odinger equation.
(ii) We compute the averaged wave function of 1000 events.
(iii) Then we compute the ground state probability.
This procedure amounts to the time evolution by $\langle H_{\rm eff}\rangle$, which is nothing but the evolution by the complex potential.

\begin{figure}[t]
\vspace{-1em}
\centering
\includegraphics[clip,width=8.5cm,bb= 0 0 380 275]{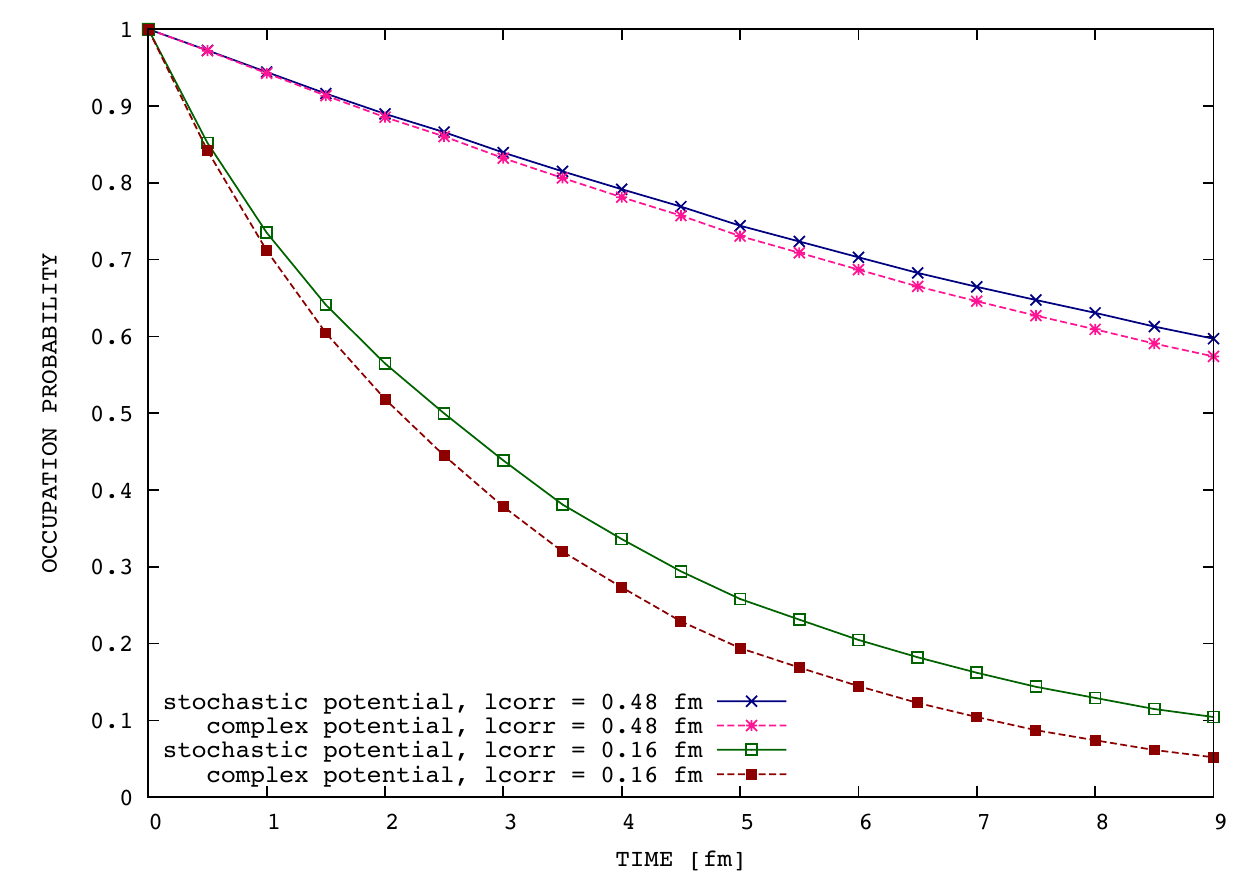}
\caption{
Time evolution of the ground state  occupation probability computed with the stochastic potential and complex potential in a static QGP.
Both the initial state and the projected state are the ground state in the Debye screened potential.
The noise correlation length in the stochastic potential is  $l_{\rm corr}=0.48, 0.16$ fm.
}
\label{fig:compare}
\end{figure}

The result is shown in Fig.~\ref{fig:compare}.
We find the fact that initially the decay rate is the same between the two approaches but the occupation at later times is different  and that a smaller $l_{corr}$ gives larger difference between two potentials. As predicted, we numerically confirm that the occupation probability is larger in the stochastic potential than in the complex potential.

\subsection{\label{subsec:level3_2}Quarkonium in a Bjorken-expanding QGP}
In relativistic heavy-ion collisions, the system rapidly expands and its temperature decreases in time.
The simplest model for such system is the boost-invariant one-dimensional expansion, or the Bjorken expansion \cite{Bjorken:1982qr}.
The temperature of a Bjorken-expanding QGP is given as a function of time $t$,
\begin{align}
T(t)=T_0 \left(\frac{t_0}{t_0+t}\right)^{1/3}.
\label{eq:T}
\end{align}
Here $T_0$ is the initial temperature and $t_0$ is the QGP formation time after heavy ions have collided.
We choose $T_0=0.4$ GeV and $t_0=1$ fm, which are typical values in relativistic heavy-ion collisions.
In this calculation, we study both bottomonium and charmonium.
The heavy quark masses and the parameters of the stochastic potential are listed in Table~\ref{table:2}.

\begin{table}
\caption{Mass and parameters in the model}
\label{table:2}
\begin{center}
\begin{tabular}{c|ccccc} \hline \hline
& $M$ [GeV] 	& \ $\alpha_{\rm eff}$ \	& $m_{D}$		& \ $\gamma$ \	& $l_{\rm corr}$  \\ \hline
Bottomonium	& 4.8					&0.3 				& $T$ 		& $0.3T$ 		& $1/T$ \\
Charmonium	& 1.18 				&0.3 				& $T$	 	& $0.3T$ 		& $1/T$ \\ \hline\hline
\end{tabular}
\end{center} 
\end{table}

\begin{figure*}
\begin{center}
\begin{tabular}{@{\hspace{-6em}}c@{\hspace{-7em}}c}
\includegraphics[clip,width=9.0cm,bb= 0 0 380 270]{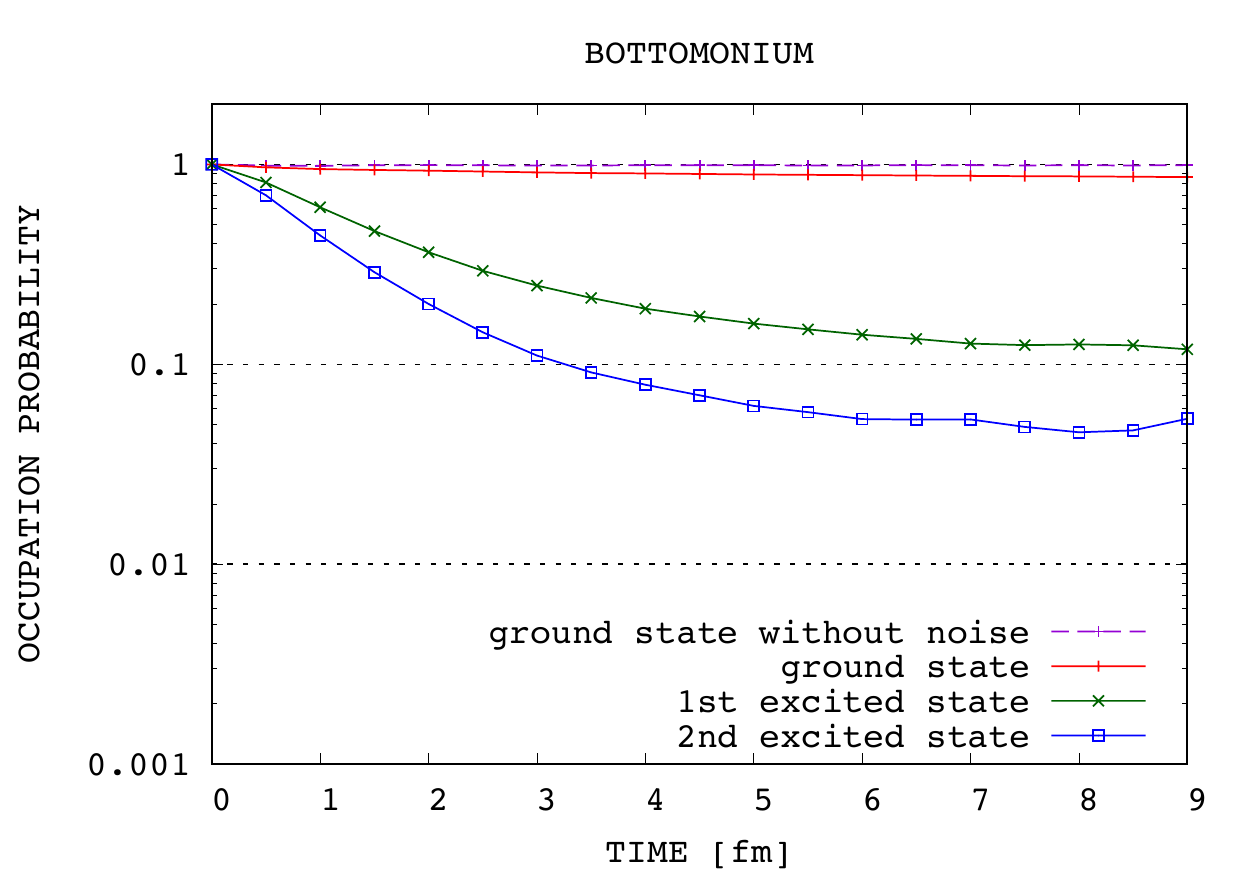}          
\includegraphics[clip,width=9.0cm,bb= 0 0 380 270]{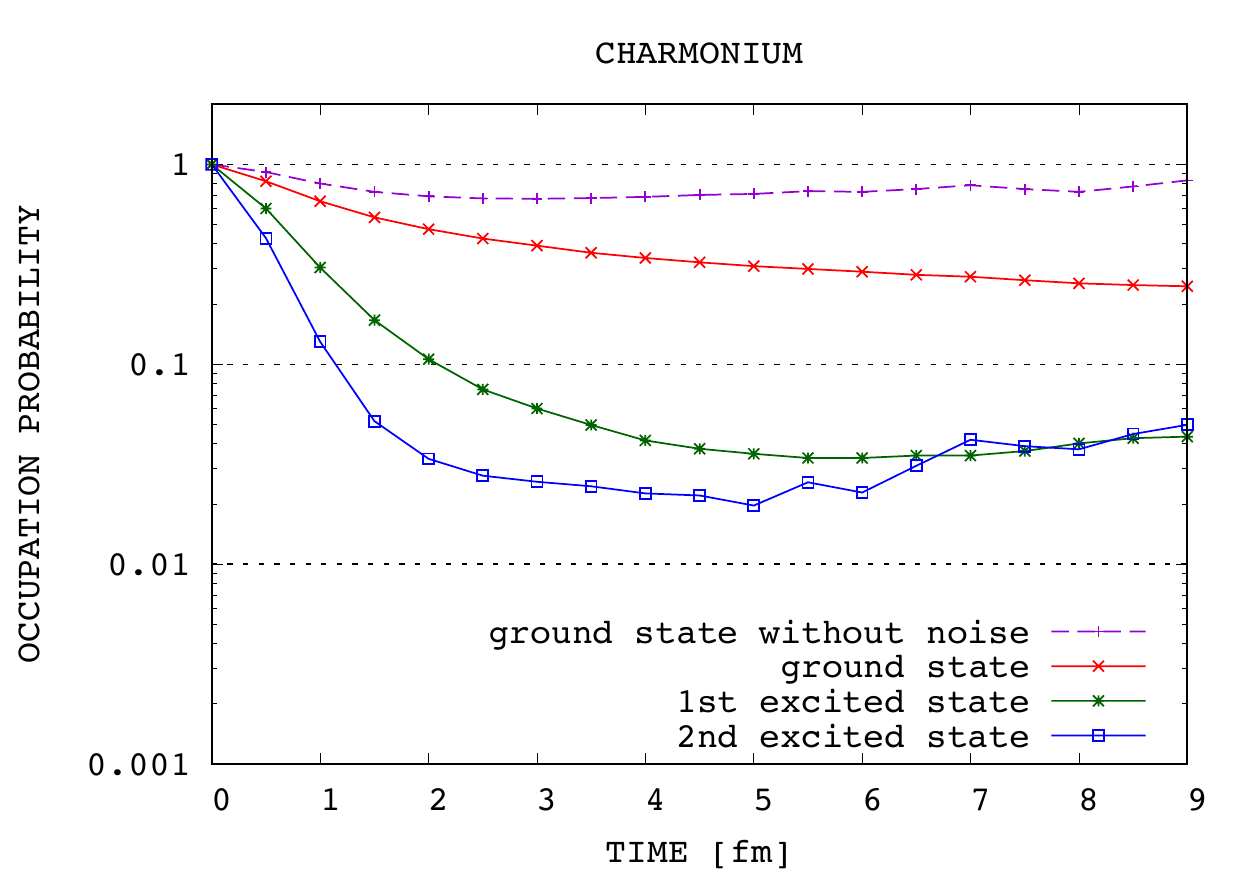}  
\\
\end{tabular}
\caption{
Time evolution of the occupation probability of quarkonium bound states (the ground, the first excited, and the second excited states) in the stochastic potential model in a Bjorken-expanding QGP.
Both the initial states and the projected states are the bound states in the vacuum Cornell potential.
The left figure shows the calculation for bottomonium and the right for charmonium.
For comparison purposes, we also plot the probability of the ground state from an evolution only with the Debye screened potential, i.e. without noise (dashed lines).
}
\label{fig:expand}
\end{center}
\end{figure*}

As the initial wave function, we chose either the ground state, the first excited state, or the second excited state of the vacuum Cornell potential,
\begin{align}
V_{\rm vac}(x) = -\frac{\alpha_{\rm eff}}{|x|} +\sigma |x|,
\label{eq:cornell}
\end{align}
with $\alpha_{\rm eff}=0.3$ and $\sigma=1$ GeV/fm. 
Here again, we regularize the singularity at the origin by $V_{\rm vac}(x)\to V_{\rm vac}(\tilde x(x))$.
The wave function is subsequently evolved by the stochastic Schr\"odinger equation.
In each event, we compute the occupation probability of the initial wave function as a function of time, then take an ensemble average over 1000 events.
In this computation, the potential in the time evolution is $V(\tilde x(x))$  [eq.(\ref{eq:stoch_pot_model})] and it depends on time while the initial and the projected states are defined from the vacuum potential $V_{\rm vac}(\tilde x(x))$.
We choose these projected states because experimentally observed particles should obey the Cornell potential in the cold medium.
Our results are shown in Fig.~\ref{fig:expand}.

We can see that the shallower bound states are excited and lost in shorter time scales within each quarkonium species. 
The reason is that shallower bound states are more extended and have larger radii, or longer coherence length $l_{\Psi}$.
Comparing $n$th states of bottomonium and charmonium, we observe that the former decreases more slowly because of their smaller radii.
These features qualitatively agree with the experimental data $R_{\rm AA}$ in relativistic heavy-ion collisions and the phenomenological expectation of sequential modification \cite{Karsch:2005nk}.

Finally, let us estimate to what extent the noise $\Theta$ is essential in the decrease of the initial bound states.
For this purpose, we start from the ground state of the Cornell potential, evolve by the Schr\"odinger equation without noise using only $U^{\langle H\rangle}_{t_n\to t_n+\Delta t}$, and compute the survival probability of the ground state.

The results are also shown in Fig.~\ref{fig:expand} (dashed lines).
The bottomonium ground state stays  almost unchanged because it is so localized that it is bound essentially in a Coulomb potential in the temperature range of our study.
The charmonium ground state occupation probability shows a nonmonotonic behavior as a function of time.
At first the Debye screened potential does not have a long-distance attractive force so that the charm and anticharm start to become separated.
At later times the temperature decreases, the Debye screening length becomes longer, and the attractive force reaches out to longer distances.
Thus the charm and anticharm are drawn to each other more closely, and the charmonium wave function becomes more localized  and has a larger overlap with the initial ground state wave function.

Let us also compare the results with and without the noise.
For both bottomonium and charmonium ground states, there is a significant deviation from the results with the stochastic potential. 
In particular for charmonium in the absence of noise, a replenishment of the ground state sets in within $t<10$fm, which is absent in the presence of wave function decoherence. 
This demonstrates explicitly that decoherence by noise represents an important dynamical mechanism for quarkonium suppression.

\section{\label{sec:level4}Conclusion}

In this paper, we studied the time evolution of a quarkonium in one-dimensional QGP using an improved stochastic potential model based on QCD.
The stochastic potential is composed of a Debye screened potential and a noise term, which possesses a finite correlation length $l_{\rm corr}$.
The scale $l_{\rm corr}$ introduces a new dimension to the quarkonium dissociation, namely wave function decoherence.
When the correlation length $l_{\rm corr}$ is much smaller than the
coherence length $l_{\Psi}$, or the radius, of a quarkonium wave function, the wave function acquires incoherent phase rotations from the noise and is easily mixed with excited states.
In the opposite case where $l_{\rm corr}\gg l_{\Psi}$, the noises for a heavy quark and a heavy antiquark nearly cancel and the wave function remains almost unaffected.
The transition of these two cases occurs at $l_{\rm corr}\sim l_{\Psi}$.
We numerically confirmed this behavior in Sec.~\ref{subsec:level3_1} for the bottomonium ground state in a static QGP.

One crucial difference between the stochastic potential compared to the complex potential is that 
 the former can define time evolution of a density matrix while the latter cannot.
This difference results in a discrepancy in the evolution of the occupation probability of initial bound states, which is calculated from the density matrix.
The probability calculated by the stochastic potential decreases more slowly than that calculated by the complex potential.
We demonstrated in Sec.~\ref{subsec:level3_1} that this discrepancy can be sizable.

We also simulated the time evolution of a bottomonium and a charmonium in a Bjorken-expanding QGP using the stochastic potential model in Sec.~\ref{subsec:level3_2}.
The initial wave functions we used were either the ground state, the first excited state, or the second excited state in the vacuum Cornell potential.
The occupation probability of the initial bound state during the evolution is computed and found to be sensitive to the bound state radius, as is expected from our study in Sec.~\ref{subsec:level3_1}.
This tendency agrees qualitatively with the results of $R_{\rm AA}$ in relativistic heavy-ion collisions.
To identify the effect of noise, we calculated the time evolution in the same setup using a potential without noise.
Again we found significant deviations between our results with stochastic potential and those only with the screened potential and thus confirmed that the wave function decoherence provides an important dynamical mechanism in the quarkonium suppression.

Our numerical simulations point out that the time evolution with the complex potential or only with the screened potential predicts unreliable survival probability of an initial bound state: The former, as depicted in Fig.~\ref{fig:compare} {\it underestimates} the occupation probability, while as in Fig.~\ref{fig:expand} the latter {\it overestimates} it.
To understand quarkonium production in relativistic heavy-ion collisions in a quantitative way with a systematic connection to QCD, wave function decoherence must be correctly treated in phenomenological studies.
%To draw a quantitative conclusion on the mechanism of quarkonium production in the relativistic heavy-ion collisions, the wave function decoherence must be correctly treated in the phenomenological models.

In the future, we extend our analysis to three-dimensional space with a more realistic evolution of the QGP and study what information about the dissociation mechanism can be learned from the quarkonium yields in the experiments.
At the same time the stochastic potential approach may be extended beyond the color singlet sector investigated here, by coupling explicit wave functions for the color octet.
In QCD with $N_c$ colors, the wave function for a heavy quark pair has color structure in $N_c \otimes N_c^*$ representation of ${\rm SU}(N_c)$ group.
The noise carries color charges $\theta^a$ ($a=1,2,\cdots, N_c^2-1$) and rotates the heavy quark colors by
%\begin{align}
$\Theta(\bm r,t) \equiv \theta^a(\bm R+\bm r/2) \left[t^a\otimes 1\right] -\theta^a(\bm R-\bm r/2) \left[1\otimes t^{a*}\right]$,
%\end{align}
with $t^a$ being the ${\bf su}(N_c)$ algebra in the fundamental representation.
An implementation of this extension, which was first discussed in \cite{Akamatsu:2014qsa}, is straightforward\footnote{
A corresponding master equation of the quarkonium states was also first derived in \cite{Akamatsu:2014qsa} and describes the transitions between color singlet and octet states.
The short distance expansion of $\mathcal D(\bm r,\bm s)$ in \cite{Akamatsu:2014qsa} has been recently reproduced by \cite{Brambilla:2016wgg}.
}.
Finally, the effect of dissipation, which has not been incorporated in our present analysis, is investigated in our future publication.

\section*{acknowledgements}

S.K., Y.A., and M.A. are thankful to Masakiyo Kitazawa for a fruitful discussion.
Y.A. thanks the German Research Foundation (DFG) Collaborative Research Centre at Heidelberg University for hospitality during his stay for ISOQUANT Mini-Workshop Quarkonium Realtime Dynamics in Heavy-Ion Collisions 2017.
A.R. acknowledges support by the DFG Collaborative Research Centre SFB 1225 (ISOQUANT).

%\begin{thebibliography}{99}  
%\end{thebibliography}

%\bibliographystyle{utphys} 
\bibliography{Ref}

\end{document}